\documentclass[twocolumn,superscriptaddress,showpacs,aps,amsmath,amssymb,prl]{revtex4}
\usepackage{bm}
\usepackage{graphicx}

\newcommand{\ep}{\epsilon}
\newcommand{\w}{\omega}
\newcommand{\W}{\Omega}

\newcommand{\be}{\begin{equation}}
\newcommand{\ee}{\end{equation}}
\newcommand{\bea}{\begin{eqnarray}}
\newcommand{\eea}{\end{eqnarray}}
\newcommand{\bsube}{\begin{subequations}}
\newcommand{\esube}{\end{subequations}}

\newcommand{\Fig}[1]{Fig.\,\ref{#1}}

\newcommand{\comments}[1]{}

\begin{document}

\title{Kondo memory in driven strongly-correlated quantum dots}

\author{Xiao Zheng} \email{xz58@ustc.edu.cn}
\affiliation{Hefei National Laboratory for Physical Sciences at the Microscale,
University of Science and Technology of China, Hefei, Anhui 230026, China}

\author{YiJing Yan} 
\affiliation{Hefei National Laboratory for Physical Sciences at the
Microscale, University of Science and Technology of China, Hefei, Anhui
230026, China}
\affiliation{Department of Chemistry, Hong Kong University of Science
and Technology, Kowloon, Hong Kong, China}

\author{Massimiliano Di Ventra} \email{diventra@physics.ucsd.edu}
\affiliation{Department of Physics, University of California, San
Diego, La Jolla, California 92093}

\date{\today}

\begin{abstract}

We investigate the real-time current response of strongly-correlated
quantum dot systems under sinusoidal driving voltages. By means of an
accurate hierarchical equations of motion approach, we demonstrate the
presence of prominent memory effects induced by the Kondo resonance on
the real-time current response. These memory effects appear as
distinctive hysteresis line shapes and self-crossing features in the
dynamic current-voltage characteristics, with concomitant excitation of
odd-number overtones. They emerge as a cooperative effect of quantum
coherence -- due to inductive behavior -- and electron correlations --
due to the Kondo resonance. We also show the suppression of memory
effects and the transition to classical behavior as a function of
temperature. All these phenomena can be observed in experiments and may
lead to novel quantum memory applications.

\end{abstract}


\maketitle


Memory of physical systems is quite a common phenomenon
\cite{Per11145}. It simply means that the state of a system at a given
time -- driven by an external input -- depends strongly on its history,
at least within a certain range of parameters of the external drive,
such as its frequency or amplitude \cite{Per11145,Yan1313}. These
memory features give rise to unconventional properties and novel
functionalities that can be used in actual device applications.

Although there has been a recent crescendo of interest in memory
effects in resistive, capacitive and inductive systems \cite{Di091717},
less work has been devoted to memory effects in strongly-correlated
quantum systems. These include the single-molecule magnets
\cite{Miy12938,Tim12104427}, metal-oxide-semiconductor thin films
\cite{Str0880,Liu12153503}, nanoparticle assemblies \cite{Kim092229},
and bulk Kondo insulators \cite{Kim12013505}. In this respect, quantum
dots (QDs) coupled to electron reservoirs are ideal platforms for
experimental and theoretical studies relevant to quantum computation
and quantum information \cite{Los98120,Ima994204}. This is because the
electron or spin state of a QD can be manipulated conveniently by
external fields, with the quantum coherence largely preserved. In
particular, at low temperatures, formation of the Kondo singlet state
\cite{Gla88452,Ng881768,Mei932601,Gol98156} opens up additional
channels for electron conduction. However, it is not at all obvious if
new features would emerge due to the interplay between quantum
coherence and electron correlations under driving conditions, and if
the Kondo phenomena would influence significantly the memory of the QD
system.

In this letter we show that Kondo resonances indeed induce prominent
memory effects when a QD is driven by an external periodic voltage.
These features appear as well-defined hysteresis line shapes and
self-crossing features in the dynamic current-voltage characteristics
with excitation of odd-number overtones. They are a direct consequence
of quantum coherence and electron correlations and are suppressed at
relatively high temperatures where a classical behavior is recovered.
All these effects can be probed at specific frequencies that are within
reach of experimental verification.

The QD system of interest is illustrated with a single-impurity Anderson
model \cite{And6141}, $H_{\rm dot} = \ep_d (\hat{n}_\uparrow + \hat{n}_\downarrow) + U
\hat{n}_\uparrow \hat{n}_\downarrow$.
Here, $\hat{n}_s = \hat{a}_s^\dag\,\hat{a}_s$, and $\hat{a}_s^\dag$
($\hat{a}_s$) creates (annihilates) a spin-$s$ electron on the dot
level of energy $\ep_d$,
while $U$ is the on-dot electron-electron (\emph{e-e}) interaction strength.
The total Hamiltonian is
\begin{equation}
H=H_{\rm dot} + H_{\rm res} +H_{\rm coupling},\label{AH}
\end{equation}
with
$H_{\rm res} = \sum_{\alpha k} \ep_{\alpha k}\,d^\dag_{\alpha k} d_{\alpha k}$
and
$H_{\rm coupling} = \sum_{\alpha k} t_{\alpha k} a^\dag_s d_{\alpha k} + {\rm H.c.}$,
for the noninteracting leads and  the dot-lead couplings, respectively.
Here, $d^\dag_{\alpha k}$ ($d_{\alpha k}$) is the
creation (annihilation) operator for the $\alpha$-lead state
$|k\rangle$ of energy $\ep_{\alpha k}$,  and $t_{\alpha k}$
is the coupling strength between the dot level and $|k\rangle$.
For numerical convenience, identical leads are considered, and their
hybridization functions assume a Lorentzian form, $\Delta_{\alpha}(\w)
\equiv \pi \sum_{k} t_{\alpha k} t^\ast_{\alpha k}\,\delta(\w -
\ep_{\alpha k}) = \Delta W^2/2[(\w - \mu_\alpha)^2 + W^2]$, where
$\Delta$ is the effective dot-lead coupling strength, $W$ is the band
width, and $\mu_\alpha$ is the chemical potential of the $\alpha$-lead.
We prepare the initial total system at equilibrium, where
$\mu_\alpha = \mu^{\rm eq} \equiv 0$. An \emph{ac} voltage is
applied to the left (L) and right (R) leads from the time $t_0
\equiv 0$, i.e., $V_{\rm L}(t) = -V_{\rm R}(t) = V_0 \sin(\W t)$,
with $V_0$ and $\W$ being the voltage amplitude and frequency,
respectively.
Upon switching on the voltage, the QD is driven out of equilibrium,
and the time-dependent current flowing into the $\alpha$-lead
$\bar{I}_\alpha(t)$ is computed.

 To proceed, a choice of method for calculating current must be made that
accurately reproduces the nonequilibrium electronic dynamics of this
strongly correlated electron system. A number of theoretical approaches
have been developed to tackle this problem, such as the time-dependent
numerical renormalization group (NRG) approach \cite{And05196801}, the
time-dependent density matrix renormalization group approach
\cite{Caz02256403,Whi04076401}, the real-time diagrammatic quantum
Monte Carlo \cite{Wer09035320,Sch09153302}, the Kadanoff--Baym
time-dependent Green's function approach \cite{Myo0867001,Thy08115333},
the time-dependent density-functional theory \cite{Kur10236801}, and
quantum master equation approaches
\cite{Tim12104427,Cui06449,Li07075114,Ped05195330}.
For instance, the time scale for the buildup of Kondo resonance in a
strongly correlated QD has been studied by using a Green's function
technique, where the \emph{e-e} interaction is
accounted for within the noncrossing approximation
\cite{Nor99808,Pli05165321}.

We choose the hierarchical
equations of motion (HEOM) approach, developed in recent years by
Yan and coworkers \cite{Jin08234703,Zhe121129,Li12266403}.
This is a unified, real-time approach
for a wide range of equilibrium and nonequilibrium,
static and dynamic properties of a
general open quantum system \cite{Zhe08184112,Zhe09124508,Li12266403,Wang2013}.
The basic variables in HEOM are the reduced density matrix of the open
system and a set of auxiliary density matrices. The HEOM
approach has been employed to study, for instance, the dynamic Coulomb
blockade and dynamic Kondo phenomena in QDs
\cite{Zhe08093016,Zhe09164708}.
To close the coupled equations, the hierarchy needs to be truncated at
a certain level $L$. In principle, the exact solution is guaranteed for
$L \rightarrow \infty$. In practice, the results usually converge
rapidly with increasing $L$ at finite temperatures. Once the
convergence is achieved, the numerical outcome is considered to be
quantitatively accurate. For instance, it has been demonstrated that
the HEOM approach achieves the same level of accuracy as the latest NRG
method for the prediction of various dynamical properties at
equilibrium \cite{Li12266403}, and it has reproduced~\cite{Zhe08184112}
the exact numerical solution for the time-dependent current response of
a noninteracting QD to a step-function bias voltage
\cite{Ste04195318,Mac06085324}.
Here, we solve the HEOM dynamics described by Hamiltonian~(\ref{AH}),
at the $L = 4$ truncation level of hierarchy
\cite{Zhe08093016,Zhe09164708} (see also Supplemental Material
\cite{SM2}). We have checked that all results presented in this letter
converge quantitatively with respect to this truncation.
%
%


\begin{figure}
\includegraphics[width=0.95\columnwidth]{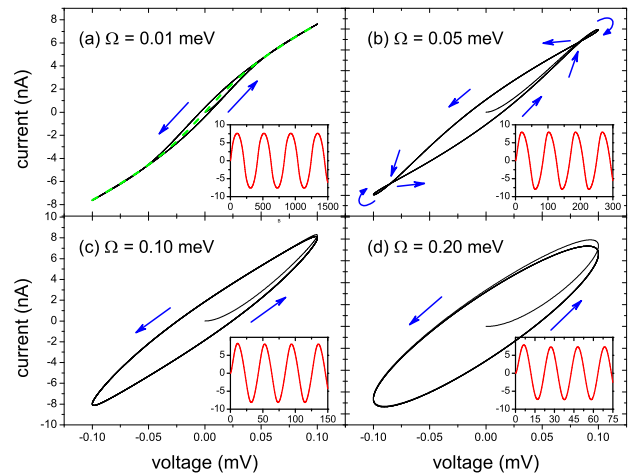}
\caption{(Color online). Current versus voltage for a QD
driven by an \emph{ac} voltage of amplitude $V_0 = 0.1\,$mV
and frequency $\Omega$.
The arrows indicate the circulating directions of the loops.
The green dashed curve in (a) depicts the steady-state $I$-$V$
characteristics. The insets plot the corresponding current (in nA) versus time (in ps).
The other parameters (in meV) are $U = -2\ep_d = 0.4$, $\Delta = 0.1$,
$W = 2$, and $T = 0.02$.
} \label{fig1}
\end{figure}

Figure~\ref{fig1} depicts the dynamic $I$-$V$ characteristics of a QD
which possesses the electron-hole symmetry ($U = -2\ep_d$) subject to
an \emph{ac} voltage of various frequencies $\Omega$. To compare with
the energetic parameters of the QD, $\Omega$ is represented by the
harmonic energy associated with the driving voltage. In particular,
$\Omega = 0.01\,$meV corresponds to a period of $0.4\,$ns. Although
high, these GHz time-domain voltage manipulations and current
measurements have been realized experimentally
\cite{Gab06499,Fev071169,Cao131401}.
The calculation results indicate that the electron-hole symmetry is
preserved under the \emph{ac} voltage, and the dot level is always
half-filled. Consequently, the displacement current is zero, and the
Kirchhoff's current law holds at any time, i.e., $I(t) = \bar{I}_{\rm
R}(t) = -\bar{I}_{\rm L}(t)$ \cite{Tim12104427}.

As shown in \Fig{fig1}(a), at a low $\Omega$, the dynamic $I$-$V$
characteristics is close to the steady-state curve, since the electrons
have sufficient time to redistribute to catch up with the variation of
voltage. In contrast, at a high $\Omega$, the dynamic $I$-$V$ curve
forms an ellipse; see \Fig{fig1}(d). The ellipse shape originates from
a phase difference between the current and voltage. In the case of
\Fig{fig1}(d), the ellipse is traversed in the counterclockwise
direction (voltage leads current), indicating a prominent inductive
behavior.
It has been demonstrated that in the linear response regime a
two-terminal device can be mapped onto a classical circuit, where a
resistor-capacitor branch and a resistor-inductor branch are connected
in parallel \cite{Yam08495203,Mo09355301,Wen115519}.
%
Here, the inductance is associated with the time for an electron to
dwell on the device \cite{Wan07155336,Yam08495203} and is
due to the fact that the dot energy is lower than the
equilibrium energy, as required in order to have a Kondo resonance.
At an intermediate $\Omega$, the dynamic $I$-$V$ curve forms a
hysteresis loop with self-crossing in the first and third quadrants;
see \Fig{fig1}(b). The hysteresis behavior highlights the nontrivial
memory effects on the real-time electron dynamics. To understand further these memory effects, we vary both the
coupling strength $U$ and the temperature $T$.


\begin{figure}
\includegraphics[width=0.95\columnwidth]{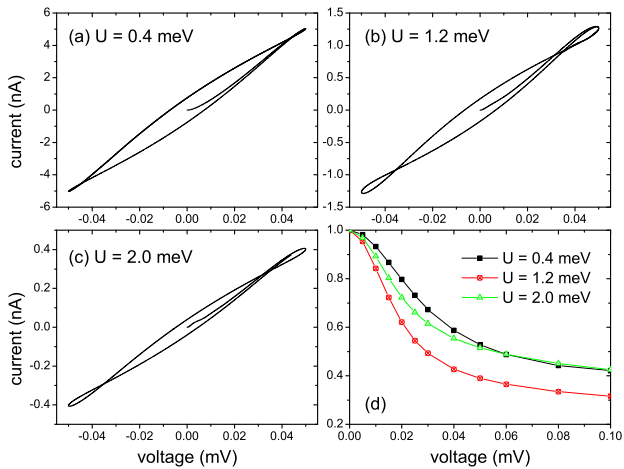}
\caption{(Color online). Current versus voltage for the QDs of
(a) $U = 0.4$, (b) $U = 1.2$, and (c) $U = 2\,$meV, under the antisymmetric
\emph{ac} voltage of $V_0 = 0.05\,$mV and $\Omega = 0.03\,$meV.
Panel (d) shows the steady-state $dI/dV$ versus the bias voltage
$V = V_{\rm L} = -V_{\rm R}$, where the displayed data are scaled
by the value of $dI/dV$ at $V = 0$.
The other parameters  (in meV) are $\ep_d = -U/2$, $\Delta = 0.1$,
$W = 2$, and $T = 0.01$.
} \label{fig2}
\end{figure}

Figure~\ref{fig2} plots the dynamic $I$-$V$ curves for a series of
symmetric QDs of different $U$ under the same driving voltage
and temperature $T$. The hysteresis behavior of \Fig{fig1}(b) is
constantly observed. It is interesting to see that the $I$-$V$ curve
shape varies in a nonmonotonic manner with respect to $U$, and the
hysteresis behavior is most outstanding at around $U = 1.2\,$meV. To
understand this trend, we calculate the steady-state differential
conductances ($dI/dV$) of the QD under the bias voltage $V_{\rm L} =
-V_{\rm R} = V$. The resulting $dI/dV$ are then scaled by their
respective values at $V = 0$. These curves are
shown in \Fig{fig2}(d). The differential conductance clearly decays nonlinearly
with increasing voltage, and $U = 1.2\,$meV corresponds to the
steepest descent of $dI/dV$.
The maximal conductance at $V = 0$ is due to the formation of Kondo
states. Under a bias voltage, the Kondo channel is blocked by the gap
between the chemical potentials in the two leads, leading to the
drastic decay in conductance with increasing $V$. This is confirmed by
the evolution of the Kondo spectral signature versus voltage (see
Supplemental Material~\cite{SM2}).
For the QD studied in \Fig{fig2}, the Kondo temperatures $T_K =
(U\Delta/2)^{1/2} \exp [ -\pi U/8\Delta + \pi\Delta /2U ]$ are $0.044$,
$2.5\times 10^{-3}$ and $1.3\times 10^{-4}\,$meV for $U = 0.4$, $1.2$
and $2\,$meV, respectively. Therefore, under a fixed temperature, the
Kondo resonance would lead to a steeper decay in $dI/dV$ with a smaller
$U$. On the other hand, the dot level of energy $\ep_d = -U/2$ becomes
closer to $\mu^{\rm eq}$ as $U$ diminishes. This provides another
conduction channel with width $\Delta$, and its associated
conductance decays much more slowly than the Kondo channel since $T_K
\ll \Delta$. With the presence of both conduction channels, the
steepest decay in $dI/dV$ occurs at an intermediate value of $U$.


The shape of the dynamic $I$-$V$ characteristics can now be interpreted
as follows. In the limit of adiabatic driving ($\Omega \rightarrow 0$),
the evolution of the current response follows exactly the steady-state
$dI/dV$-$V$ curve. The reduction in $dI/dV$ versus $V$ leads to a
concave $I$-$V$ curve at $V
> 0$; see the green dotted line in \Fig{fig1}(a). At a finite $\Omega$,
the inductive feature starts to emerge, which forces the current to lag
behind the driving voltage. For instance, at an instant when the
voltage turns from negative to positive, the current has not yet
reached its turning point -- it remains negative but accelerates toward
the voltage. Consequently, the dynamic $I$-$V$ curve has a convex
curvature at $V = 0$ and $I < 0$. The convexity may be inverted into
concavity provided that the increase in voltage captures the drastic
decay in conductance. This requires a relatively low $\Omega$, with
which the variation of ``transient'' conductance does not deviate much
from the steady-state $dI/dV$-$V$ curve shown in \Fig{fig2}(d). Such a
convex-concave curve segment, together with its mirror-image
counterpart upon voltage decrease, results in a self-crossing
hysteresis loop as depicted in \Fig{fig1}(b) and \Fig{fig2}(a)-(c).
At a high $\Omega$, there is not enough time for the buildup of Kondo
states, and thus the $dI/dV$-$V$ curve is effectively smoothed. In this
case, the inductive feature becomes a dominant factor, and the dynamic
$I$-$V$ curve displays a simple ellipse shape; see \Fig{fig1}(d).

%
%
%

Based on the above analysis, we can conclude that the hysteresis behavior of the \emph{ac}
$I$-$V$ characteristics originates from a cooperative interplay between
the quantum coherence (the inductive feature) and strong electron
correlation (the Kondo resonance). Since both quantum coherence and
correlation are significantly influenced by thermal fluctuations, the
associated memory effects are expected to depend sensitively on the
temperature.

\begin{figure}
\includegraphics[width=1.0\columnwidth]{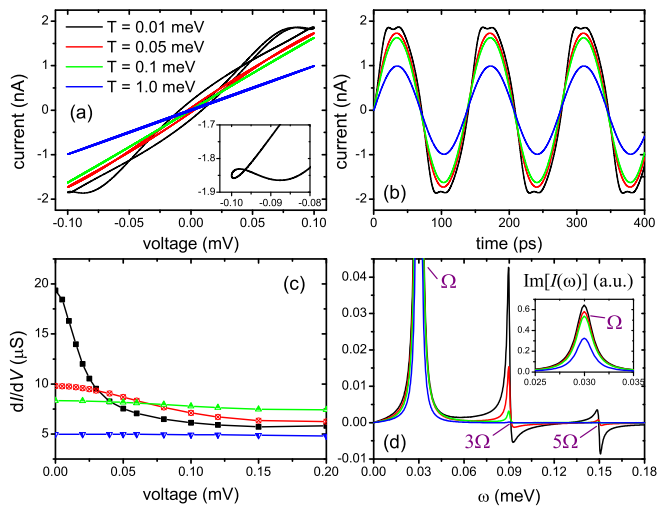}
\caption{(Color online). (a) Dynamic $I$-$V$ characteristics,
(b) real-time current response, (c) steady-state $dI/dV$ versus $V$,
and (d) imaginary part of current spectrum
of a QD with $U = -2\ep_d = 1.2\,$meV at various temperatures.
The inset of (a) magnifies the self-crossing of \emph{ac} $I$-$V$ curve
near $V = -V_0$ at $T = 0.01\,$meV, and
that of (b) displays the full peaks of ${\rm Im}[I(\w)]$ at
$\w = \Omega$.
The other parameters  (in meV) are $\Delta = 0.1$,
$W = 2$, $V_0 = 0.1$, and $\Omega = 0.03$.
} \label{fig3}
\end{figure}

In \Fig{fig3}(a) we show the evolution of the dynamic $I$-$V$ curve
with increasing temperature. Here, in order to accentuate the nonlinear
current response, a somewhat larger voltage amplitude ($V_0 =
0.1\,$meV) is adopted. At low temperature ($T = 0.01\,$meV), the
\emph{ac} $I$-$V$ curve displays a remarkable hysteresis behavior, with
the emergence of multiple self-crossing points. Such a convoluted line
shape is a strong indication of complex memory effects. With the
temperature rising by as little as $0.01\,$meV, the hysteresis behavior
of \Fig{fig2}(b) is recovered, with only one crossing in the first or
third quadrant.
%
This apparent change in the hysteresis pattern should be ascribed to
the suppression of the Kondo resonance. Indeed, as shown clearly in \Fig{fig3}(c),
the zero-bias conductance decays drastically with a minor increase in
$T$ from $0.01\,$meV.
Meanwhile, increasing $T$ leads to the progressive narrowing of the
$I$-$V$ hysteresis loop, and the loop almost merges into a single line
at $T = 0.1\,$meV. This means that the inductive feature becomes
irrelevant at a high temperature, and the QD behaves like a classical
resistor. This is because the conducting electrons have a wide energy
distribution due to thermal excitations, and the phase difference
between the current and voltage is largely randomized. Due to the same
reason, the steady-state $dI/dV$ varies rather smoothly as a function
of $V$ at high $T$, leading to an overall Ohmic conductance.


The temperature-sensitive memory is further explored through a
frequency analysis of the response current. The current spectrum for a
single-lead QD under sinusoidal voltages has been studied in
Ref.\,\onlinecite{Mo09355301}, where the response current signals
concentrate on the exact multiples of driving frequency $\Omega$. For
the two-terminal QD investigated here, the variation of current
spectrum versus temperature is depicted in \Fig{fig3}(d). It is found
that the response current arises predominantly at $\w = (2n+1)\Omega$
with $n \in \mathbb{Z}$, while the even overtones remain unexcited.
This is because the \emph{ac} voltages have the left-right
antisymmetry, so that the response current would flow in the reverse
direction if the half-period were taken as the initial time ($t_0 =
\pi/\W$). Therefore, the relation $I(\w) = -I(\w) e^{i\w\pi/\W}$ must
be satisfied. Overtones of all symmetries can be found by breaking the
symmetric coupling between the QD and the leads \cite{Coh12133109}. As
$T$ rises from $0.01$ to $0.1\,$meV, the change in current amplitude at
the fundamental frequency is rather minor. In contrast, the overtone
responses vanish almost completely. This highlights the dynamic
Kondo-assisted electron conduction, which requires a relatively low
excitation energy. At $T < 0.01\,$meV, the Kondo memory effects are
expected to be even more prominent. However, to achieve convergent HEOM
results at a lower $T$, a higher truncation level $L$ is required,
which is numerically too demanding with the present computational
resources at our disposal.
At $T > 0.1\,$meV, the Kondo resonance is absent, and the dot level is
away from the resonance region ($|\ep_d|
> V_0$). Consequently, the system falls in the linear response regime,
and the corresponding real-time current becomes synchronized with the
voltage; see \Fig{fig3}(b).
%


\begin{figure}
\includegraphics[width=0.85\columnwidth]{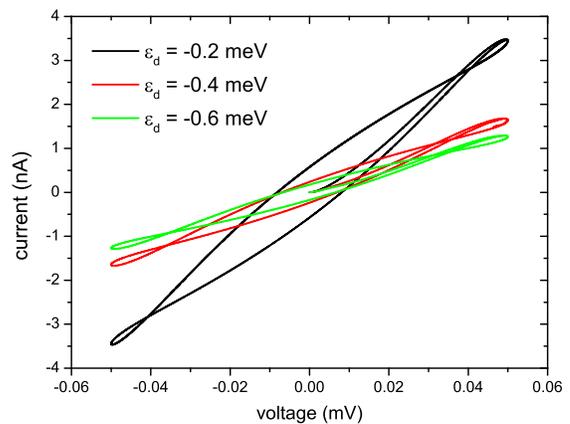}
\caption{(Color online). Dynamic $I$-$V$ characteristics
of QDs with different $\ep_d$. The displayed current is
$I(t) = [\bar{I}_{\rm R}(t) - \bar{I}_{\rm L}(t)]/2$.
The other parameters (in meV) are $\Delta = 0.1$, $U = 1.2$, $T = 0.01$,
$W = 2$, $V_0 = 0.05$, and $\Omega = 0.03$.
} \label{fig4}
\end{figure}

We finally present some considerations on the experimental verification
of our predictions. In realistic experimental setups, a QD may involve
more than one energy level and more complicated forms of \emph{e-e}
interaction. Therefore, the memory effects will influence the real-time
dynamics in a rather complex way. However, since the Kondo resonance is
tightly tied to the lead chemical potential, the Kondo memory may be
identified distinctly from other memory effects. For instance, consider
the scenario where the dot-level $\ep_d$ is shifted by a gate voltage,
and the \emph{ac} $I$-$V$ curves at different $\ep_d$ are shown in
\Fig{fig4}. At $U \neq -2\ep_d$, the electron-hole symmetry is broken,
and transient charging takes place at the QD.
To concentrate on the net conduction current, the displacement
component is taken care of by symmetrizing the left- and right-lead
currents~\cite{Tim12104427}. As clearly indicated in \Fig{fig4}, the
Kondo-resonance-induced hysteresis behavior is continually preserved
with $\ep_d$ shifted by as much as $0.4\,$meV. Since any
single-electron resonance associated with a dot energy level would have
been switched off under such a large gate voltage, \Fig{fig4}
accentuates the robustness of Kondo memory effects.

%
%

To conclude, we have demonstrated the presence of Kondo-resonance-induced
memory effects in the real-time electronic dynamics of strongly
correlated QDs, which may be observed in experiments.
The predicted hysteresis line shape and self-crossing features of the
\emph{ac} $I$-$V$ characteristics highlight the remarkable interplay
between quantum coherence and quantum electron correlations.
These effects may lead to novel applications of QDs in the Kondo regime,
ranging from machine learning \cite{Per09021926} to massively-parallel computation
and information processing \cite{Per11046703}.
%

%


%

%

%
The support from the NSF of China (No.\,21103157, No.\,21233007, and
No.\,21033008) (XZ and YJY), the Fundamental Research Funds for Central
Universities (No.\,2340000034 and No.\,2340000025) (XZ), and the Hong
Kong UGC (AoE/P-04/08-2) and RGC (No.\,605012), is gratefully
acknowledged. MD acknowledges support from the NSF grant
No.\,DMR-0802830 and the Center for Magnetic Recording Research at
UCSD.
%






\end{document}